\documentstyle[12pt]{article}
\begin{document}
\begin{center}
{\large{\bf Liquid-gas phase transition in nuclei in the 
relativistic  Thomas-Fermi theory}}\\
\vspace*{0.20  true in}
Tapas Sil$^1$\footnote{e-mail: tapassil@veccal.ernet.in},
B. K. Agrawal$^2$\footnote{e-mail: bijay@tnp.saha.ernet.in}, 
 J. N. De$^1$\footnote{e-mail: jadu@veccal.ernet.in} and
S. K. Samaddar$^2$\footnote{e-mail: samaddar@tnp.saha.ernet.in}\\
%\vspace*{0.20 true in}
$^1$ Variable Energy Cyclotron Centre, \\
1/AF Bidhannagar, Calcutta 700064, India\\
$^2$ Saha Institute of Nuclear Physics, \\
1/AF Bidhannagar, Calcutta 700064, India\\
\end{center}

\begin{abstract}
The equation of state (EOS) of finite nuclei is constructed in
the relativistic Thomas-Fermi theory using the non-linear $\sigma-\omega
-\rho$ model.  The caloric curves are calculated by confining 
the nuclei in the freeze-out volume taken to be a sphere of size
about 4 to 8 times the normal nuclear volume.
The results obtained from the relativistic theory  are not 
significantly different from those
obtained earlier in a non-relativistic framework. 
The nature of the EOS and the peaked 
structure of the specific heat $C_v$ obtained from the caloric curves show
clear signals  of a liquid-gas phase transition in finite nuclei.
The temperature evolution of the Gibbs potential and the entropy at
constant pressure indicate that the characteristics of the transition 
are not too different from the first-order one.
\end{abstract}

\vskip 0.5cm
PACS number(s): 25.70.Pq, 21.10.-K, 21.60.-n, 21.65.+f

\newpage
\section{Introduction}
Nuclear caloric curves have been obtained from a number of
experiments in recent times from energetic nucleus-nucleus collisions. 
In the GSI data \cite{poc} for $Au+Au$ at 
600 $A$MeV, the temperature was found to be practically constant at
a value of $T\sim 5$ MeV over the excitation
energy ($\epsilon^*$) range of 3 - 10 MeV per nucleon
after which $\epsilon^*$ was found to increase  linearly
with temperature as in a classical gas. This is 
suggestive of a sharp liquid-gas type phase transition. 
The caloric curve obtained from the collision of  $Au$ on $C$ at 1 $A$GeV in
the EOS collaboration experiment \cite{hau} also shows
a plateau at $T\sim 6$ MeV; this is not as prominent
as in the earlier case. However, the heat capacity
$C_v$ derived from  this caloric curve of the EOS group 
shows a peaked structure at $T\sim 6$ MeV indicating  existence of  a
phase transition.  Even at a  relatively low
bombarding energy of 47 $A$MeV \cite{cib} for several reactions, it has
been seen recently that the caloric curves show plateau at
$T\sim 7$ MeV in the excitation energy range of
$\sim 3.5$ to $7$ MeV per nucleon. It would
be interesting to know whether this is a  precursor  to the  liquid-gas
phase transition in the finite nuclei. Theoretical analysis of  infinite nuclear matter
(symmetric as well as asymmetric), both in the non-relativistic
\cite{jaq,ban} and relativistic framework \cite{ser,mul} predict 
Van der Waals type isotherms in their equation of state
(EOS) implying  coexistence of liquid and gas phases.
Finite size effects and the Coulomb interaction between protons
might change   such a behavior,  but the EOS of realistic nuclei,
calculated  only recently \cite{de1} in the non-relativistic 
Thomas-Fermi (TF) theory display same kind of isotherms, liquid-gas
coexistence  and a liquid-gas phase transition at a temperature
quite below the critical temperature for infinite nuclear matter.
The calculated transition temperatures are still somewhat higher
than the observed ones, the caloric curves do not
match exactly  the ones derived from  experiments, but
the calculations in Ref.\cite{de2} give unmistakable signals   
of the liquid-gas
phase transition in the finite size nuclear systems.

Relativistic mean field (RMF) theories have been applied
successfully  to explain  the ground state properties of
nuclei over the entire periodic table \cite{gam}.
This theory  has also  proved to be very  fruitful in explaining various
details of exotic nuclei near the drip  lines \cite{lal,vre}.
In contrast to the non-relativistic models, the RMF theory uses a single set
of parameters to explain all these properties. It
would therefore be very  interesting to investigate  the EOS of finite nuclei
and the  related exotic phenomena like liquid-gas phase
transition in the relativistic approach. The present
paper aims at understanding these thermodynamic properties of hot nuclei in a 
relativistic Thomas-Fermi (RTF) theory.

In Sec. 2, we briefly outline the formalism used.
The results and discussions are presented in Sec. 3.
The summary and conclusions are given in Sec. 4.
\section{Formalism}
A brief outline of the calculations of the relevant
thermodynamic quantities in the
relativistic Thomas Fermi approximation is presented in this section. 
The  Lagrangian density  used is given by \cite{gam}
\begin{eqnarray}
\label{lag}
{\cal L}&=& \bar\Psi_i\left ( i\gamma^\mu \partial_\mu - M\right )\Psi_i
+ \frac{1}{2} \partial^\mu\sigma\partial_\mu\sigma - U(\sigma)
- g_\sigma \bar\Psi_i \sigma\Psi_i\nonumber\\
&& - \frac{1}{4}\Omega^{\mu\nu}\Omega_{\mu\nu}
+\frac{1}{2}m_\omega^2\omega^\mu \omega_\mu - g_\omega \bar\Psi_i \gamma^\mu
\omega_\mu\Psi_i 
-\frac{1}{4}\vec{R}^{\mu\nu} \vec{R}_{\mu\nu} + \frac{1}{2} m_\rho^2 \vec{\rho}^\mu\vec{\rho}_\mu\nonumber\\
&&- g_\rho \bar\Psi_i \gamma^\mu\vec{\rho}_\mu\vec{\tau}\Psi_i
-\frac{1}{4}F^{\mu\nu}F_{\mu\nu} - e\bar\Psi_i \gamma^\mu \frac{(1-\tau_3)}{2} A_\mu\Psi_i
.
\label{Lag}
\end{eqnarray}
The meson fields included are those of the isoscalar $\sigma$ meson, 
the isoscalar-vector $\omega$
meson and the isovector-vector $\rho$ meson. The arrows in Eq. (\ref{Lag})
denote the isovector quantities. The z-component of isospin, $\tau_3$,
is taken to be $+1$ for neutrons and $-1$ for protons.  
For appropriate description of the nuclear surface properties \cite{bog},
scalar self-interaction term $U(\sigma)$ of the $\sigma$ meson
is included in the Lagrangian,
\begin{equation}
U(\sigma) = \frac{1}{2}m_\sigma^2 \sigma^2 + \frac{1}{3}g_2 \sigma^3 + 
\frac{1}{4}g_3\sigma^4
.
\end{equation}
The quantities 
$M$, $m_\sigma$, $m_\omega$ and $m_\rho$ are the nucleon, $\sigma$, 
$\omega$ and the $\rho-$meson masses, respectively, while 
$g_\sigma$, $g_\omega$, $g_\rho$ and $e^2/4\pi = 1/137$ are
the corresponding coupling constants for the mesons and the photon. 
The field tensors of the
vector mesons and of the electromagnetic fields have the following structure:
\begin{eqnarray}
\Omega^{\mu\nu} & = & \partial^\mu\omega ^\nu - \partial^\nu\omega^\mu,\\
{\vec {\bf R}}^{\mu\nu} & = & \partial^\mu {\bf{\vec \rho}}^\nu 
- \partial^\nu{\bf {\vec \rho}}^\mu - g_\rho(\vec\rho^\mu\times\vec\rho^\nu),\\
F^{\mu\nu}& = & \partial^\mu A^\nu - \partial^\nu A^\mu.
\end{eqnarray}

The equations of motion are obtained from the variational principle. 
The mean field approximation is introduced at this stage by treating the fields as $c-$numbers or classical
fields. This results  in a set of coupled equations, namely the Dirac equation with potential
terms for the nucleons and the Klein-Gordon type equations with sources  for the
mesons and the photon. 
Time reversal invariance and charge conservation get the equations simplified
in the static case. The resulting equations, known as relativistic mean-field
equations, have the following form.
The Dirac equation  for the nucleon is  
\begin{equation}
\label{dir}
\{-i{\bf\alpha}\cdot{\bf \nabla} + {\cal V}(\tau_3,{\bf r}) 
+\beta\left [M+{\cal S}({\bf r})
\right ] \} \Psi_i = \epsilon_i\Psi_i,
\end{equation}
where
\begin{equation}
{\cal V}(\tau_3,{\bf r}) = g_\omega \omega_0({\bf r})  
+ g_\rho \tau_3{\bf \rho}_0({\bf r})
+ e\frac{(1 - \tau_3)}{2} A_0({\bf r}),
\end{equation}
and 
\begin{equation}
{\cal S}({\bf r}) = g_\sigma\sigma({\bf r}),
\end{equation}
are  the {\it vector } and the {\it scalar} potentials, respectively. 
The {\it scalar} potential contributes  to the effective mass as 
\begin{equation}
M^*({\bf r}) = M + {\cal S}({\bf r})
.
\end{equation}

The Klein-Gordon equations for the mesons and the electromagnetic fields  with the
nucleon densities as sources  are 
\begin{eqnarray}
\label{sig}
\left \{ -\Delta + m_\sigma^2\right \} \sigma({\bf r}) & = & -g_\sigma\rho_s({\bf r})
-g_2\sigma^2({\bf r}) - g_3\sigma^3({\bf r}),\\
\label{ome}
\left \{ -\Delta + m_\omega^2\right \} \omega_0({\bf r}) 
& = & g_\omega \rho_v({\bf r}),\\
\label{rho}
\left \{-\Delta +  m_\rho^2\right \} \rho_0({\bf r})  & = & g_\rho\rho_3({\bf r}),\\
\label{pho}
-\Delta A_0({\bf r}) & = & e\rho_c({\bf r}).
\end{eqnarray}
While considering a finite nucleus, for simplicity, we assume it to be
spherically symmetric.
The above field equations then
can be written in a general form 
\begin{equation}
\left ( -\frac{\partial^2}{\partial r^2} - \frac{2}{r}\frac{\partial}{\partial r}
+m_\phi^2\right )\phi(r)=S_\phi(r),
\label{phi}
\end{equation}
$m_\phi$ are the meson masses for $\phi = \sigma, \omega$ and $\rho$
and is zero for the  photon. The source
 term $S_\phi(r)$  is given by the right hand side of  the 
Eqs. (\ref{sig}) - (\ref{pho}) for $\sigma, \omega, \rho$ and Coulomb fields.
The above equation (\ref{phi})  can be solved using Green's function \cite{gam}
\begin{equation}
\phi(r)= \int_0^\infty r^{'2} G_\phi(r,r') S_\phi(r') dr',
\end{equation}
where
\begin{equation}
G_\phi(r,r')=\frac{1}{2m_\phi r r'}[e^{-m_\phi\mid r - r'\mid } - e^{-m_\phi\mid 
r + r'\mid}] ,
\end{equation}
 for the massive fields and
\begin{eqnarray}
G_\phi(r,r')=\left \{
\begin{array}{cc}
\frac{1}{r}& \qquad {\rm  for \> \> r > r'}\\
\frac{1}{r'}& \qquad {\rm  for \> \>  r<  r'} ,
\end{array}
\right .
\end{eqnarray}
for the Coulomb field.

The quantities $\rho_s$, $\rho_v$, $\rho_3$ and $\rho_c$   
appearing on the right hand side of Eqs. 
(\ref{sig}) - (\ref{pho}) are the scalar, baryon, isovector and
charge densities, respectively. They can be obtained as 
\begin{eqnarray}
\rho_s(r)&=& \sum_{\tau_3}\rho_s(\tau_3,r) ,\nonumber\\
\rho_v(r)&=& \sum_{\tau_3}\rho_v(\tau_3,r) ,\nonumber\\
\rho_3(r)&=& \sum_{\tau_3}\tau_3\rho_v(\tau_3,r) ,\nonumber\\
\rho_c(r)&=& \sum_{\tau_3}\left (\frac{1-\tau_3}{2}\right )\rho_v(\tau_3,r)
.
\end{eqnarray}
In the Thomas-Fermi approximation, 
the quantities $\rho_v(\tau_3,r)$ and $\rho_s(\tau_3,r)$ are
given as
\begin{eqnarray}
\rho_v(\tau_3,r)&=& \frac{\gamma}{2\pi^2}\int_0^\infty f(\epsilon,T)k^2 dk ,\\
\rho_s(\tau_3,r)&=& \frac{\gamma}{2\pi^2}\int_0^\infty 
\frac{M^*(r)}{\sqrt{k^2+M^{*2}(r)}}f(\epsilon,T)k^2 dk,
\end{eqnarray}
where the spin degenaracy factor $\gamma $ 
is equal to 2. The self-consistent occupancy function $f(\epsilon , T)$
is obtained through the minimisation of the thermodynamic potential
\begin{eqnarray}
G= E-TS -\mu N,
\end{eqnarray}
and is given by  
\begin{equation}
f(\epsilon,T)=\frac{1}{1+e^{(\epsilon-\mu)/T}} ,
\end{equation}
with
\begin{equation}
\epsilon(\tau_3,k,r)={\cal V}(\tau_3,r)+\sqrt{k^2+M^{*2}(r)}.
\end{equation}
The chemical potential $\mu$ is adjusted to get the desired number of
particles (neutrons and protons) given by
\begin{equation}
n(\tau_3)=4\pi \int_0^{r_{max}} r^2 \rho_v(\tau_3,r) dr.
\end{equation}
In Eq.(24), $r_{max}$ determines the confining volume $V$ taken to be spherical.
From Eqs.(19) and (22), it can be seen that at large distances the baryon 
density $\rho_v$ $\sim$ $e^{(\mu-M)/T}$ and is therefore a non-zero constant at 
finite temperature. The solution to baryon density and hence the various 
observables depend on the choice of the size of the box in which the calculation
is performed. At zero temperature, however, the solution is independent of
the choice of the confining volume once it is larger than the normal nuclear
volume $V_0$. Exactly the same characteristic is seen in the nonrelativistic
case \cite{bra,de3}. The choice of the volume for the evaluation of the thermodynamic
variables would be discussed in the next section.

For a nuclear system with mass number $A$, the total energy $E(T)$ is given by
\cite{gam},
\begin{equation}
\label{ene}
E(T) = E_{part} + E_\sigma + E_{\sigma NL}+E_\omega + E_\rho 
+ E_C + E_{CM} - AM,
\end{equation}
with
\begin{eqnarray}
\label{epart}
E_{part} &=& \frac{2\gamma}{\pi}\sum_{\tau_3}
\int_0^{r_{max}} r^2 dr\int_0^\infty k^2 \epsilon(\tau_3,k,r)
f(\epsilon,T) dk ,\\
\label{esi}
E_\sigma& = &-\frac{1}{2}g_\sigma \int d^3r \rho_s({\bf r}) 
\sigma({\bf r}),\\
\label{esinl}
E_{\sigma NL}& = &-\frac{1}{2} \int d^3r
\left\{\frac{1}{3}g_2 \sigma^3({\bf r})+\frac{1}
{2}g_3\sigma^4({\bf r})\right\},\\
\label{emo}
E_\omega& = &-\frac{1}{2}g_\omega\int d^3r \rho_v({\bf r})\omega^0({\bf r}),\\
\label{erh}
E_\rho& = &-\frac{1}{2}g_\rho\int d^3 r \rho_3({\bf r}) \rho^0({\bf r}),\\
\label{ecou}
E_C& = &-\frac{e^2}{8\pi}\int d^3r \rho_C({\bf r}) A^0({\bf r}),\\
\label{ecm}
E_{CM}& = &-\frac{3}{4}\hbar\omega_0 = -\frac{3}{4}41 A^{-1/3}.
\end{eqnarray}

The free energy $F$ is given by $(E-TS)$ where 
the entropy $S$ can be calculated from the Landau quasiparticle approximation, 
\begin{equation}
\label{entropy}
S = -\frac{2\gamma}{\pi}\sum_{\tau_3}\int_0^{r_{max}}r^2 dr
\int_0^\infty k^2\left [ flnf+(1-f)ln(1-f)\right ] dk.
\end{equation}
The specific heat , $C_v$, and the pressure, $P$, can be
calculated from  
\begin{eqnarray}
\label{cv}
C_v&=& \left . \frac{dE}{dT} \right |_V ,\\
P&=& \left . -\frac{dF}{dV} \right |_T .
\end{eqnarray}
The baryonic density, the mesonic and the Coulomb fields  are obtained
iteratively through the following scheme:\\
i) An initial guess is made for the fields $\sigma(r), \omega(r),\rho(r)$
and $A_0(r)$.\\
ii) The effective mass $M^{*}$ and the energy $\epsilon$ given by Eqs.
(9) and (23), respectively are calculated with the guess fields. The
proton and the neutron chemical potentials ($\mu$) are adjusted to
reproduce the given number of nucleons of each kind.\\
iii) From Eq.(18), the various densities and hence the source terms
are calculated.\\
iv) These source terms are used in Eqs. (10)- (13) for the generation of the
new fields.

The steps (ii)- (iv) are repeated till the desired accuracy is reached.

The expressions for the EOS for nuclear matter can be  obtained
as a special case of a finite nucleus by ignoring the gradient terms
in the field equations. This simplifies the expressions for the
relevant observables and are given in Ref.\cite{ser}.

\section{Results and discussions}

In this section, the results of our calculations for the 
EOS of infinite symmetric and asymmetric nuclear matter as well as
of a few finite nuclear systems are first given. We have 
chosen $^{40}Ca$, $^{109}Ag$ and $^{150}Sm$ 
as the reprersentative systems. The results of our calculations
of the caloric curve for these nuclei are next presented. A host of
parameter sets for the nonlinear $\sigma -\omega -\rho$ model are
available which produce almost similar ground state properties
of nuclei over the whole periodic table but with widely different
values of nuclear incompressibility ($K_\infty$). To study the effects of
different $K_\infty$ on the results of our calculations, we have chosen
the parameter sets NL1, NL3 and NLSH \cite{gam,lal2} having $K_\infty$ equal
to 212, 272 and 356 MeV, respectively. 

\subsection{The nuclear EOS}
The EOS of symmetric and asymmetric nuclear matter in the RTF theory
has already been calculated by M$\ddot u$ller and Serot \cite{mul}. For
comparison of results for finite nuclei obtained with a given set of
parameters and also for completeness, we have repeated the calculations
for infinite systems with the same given parameter set. 
In Fig.1, the isotherms of symmetric nuclear matter (top panel) and 
asymmetric nuclear matter (bottom panel) with $X$ =0.2 are displayed.
The asymmetry parameter is defined as $X= (\rho_n -\rho_p) /
(\rho_n +\rho_p) $. 
The abscissa refers to $\rho_0/\rho$ ($=V/V_0$) where $\rho_0$ is the 
saturation density of normal nuclear matter given by $0.15$ fm$^{-3}$.
The isotherms are calculated with the NL3 parameter
set and are shown for a few temperatures, at and around 
the critical temperature $T_c$. For
symmetric matter, $T_c$ is 14.2 MeV whereas for 
the asymmetric matter considered,
it is 13.6 MeV. The isotherms resemble closely  those obtained for the 
Van der Waals systems and are not quantitatively very different from
those found in the nonrelativistic approach \cite{de1}. The dashed
and the dotted lines in the figure are the liquid-gas coexistence line
and the spinodal
line, respectively. With the other parameter sets, the results are very
similar; however, the critical temperature is found to increase and
the critical volume becomes smaller for
the parameter set which gives larger incompressibility. The critical
temperature and the critical volume for the different parameter sets are
given in Table I.

For the symmetric infinite system, the liquid-gas coexistence line is obtained
from the Maxwell construction as it is effectively a one-component system.
Here the pressure and the chemical potential 
in the two phases are the same at a fixed
temperature (Gibbs criteria) throughout their coexistence for any
value of $\lambda$ where $\lambda$ is the liquid volume fraction
(the gas volume fraction is $1-\lambda$). Then the neutron-proton
asymmetry is zero in both phases for all values of $\lambda$. The
asymmetric nuclear matter is a two-component system; here for any 
$\lambda$, not only that the Gibbs criteria 
are to be fulfilled for thermodynamical
coexistence of the liquid and gas phases, the overall asymmetry 
(neutron-proton ratio) has to be conserved which introduces added
complications. Now, the pressure, chemical potentials  and the 
neutron-proton ratio in both phases are in general changing functions 
of the volume fraction \cite{mul}. We have taken these factors into
account to determine the liquid-gas coexistence region. The spinodals
shown in Fig.1 are the isothermal spinodals referring to mechanical
instability. The diffusive spinodal \cite{bar} for the asymmetric matter
is not shown here.

Finite nuclei (even symmetric ones) with Coulomb interaction switched on
behave like two-component systems. However, the ideas expounded earlier
for asymmetric nuclear matter for the construction of the liquid-gas
coexistence lines can not be employed to the case of finite systems.
The density in finite nuclei is not uniform unlike infinite nuclear matter
which is homogeneous in either the liquid or the gas phase. For asymmetric
infinite system, we have found that for the liquid-gas coexistence, in
general the neutron-proton ratio in the gas phase is much larger than
that in the liquid phase. There is thus a phase separation between the
neutrons and protons. For the construction of the isotherms for finite
nuclei, the system is enclosed in a finite volume; then the self-consistent
solution of the density profiles does not allow any significant
neutron-proton phase separation  because of the strong attractive
unlike pair interactions compared to the interaction among the like
pairs. Indeed, from numerical calculations, we find that the $N/Z$ 
ratio throughout the nuclear volume is nearly the same (not varying
by more than 10\%) at all temperatures beyond $T \sim$ 3.0 MeV. For
a two-component thermodynamic system, liquid-gas coexistence occurs 
along the Maxwell-line (constant pressure, equal areas in the unstable
phases in the $P-V$ diagram) if the ratio of the concentrations of
the components are the same in the two phases as it is then 
effectively a one-component system. It is known that for  small
finite one-component systems, the pressure in the coexistence region
may have a small negative slope \cite{pat} in the $P-V$ plane, but for
symmetric nuclei with Coulomb switched off, we find numerically that 
the conventional Maxwell construction is an excellent approximation
as the differences in the chemical potentials on both ends of the
Maxwell line are negligibly small.
It is found that for the asymmetric finite nuclei under study (even with Coulomb
on), the difference in the neutron or the proton chemical potentials 
at the ends of the Maxwell line is typically $\sim$ 0.2 MeV only which
is around 30-40 times smaller compared to that for infinite nuclear matter
with the same asymmetry.
We therefore expect that for these finite systems,
for the determination of the liquid-gas coexistence,
conventional Maxwell construction may not be a poor approximation 
to which we have resorted to in the present calculations for the
nuclei considered.

The isotherms for the lightest nucleus $^{40}Ca$ and
the heaviest nucleus $^{150}Sm$ that we consider are shown in 
figures 2 and 3, respectively  for the parameter set NL3.
The results with  the other parameter sets are not displayed as they
look very similar. The finite size effects and the Coulomb interactions
between protons do not change the qualitative character of the 
isotherms, the only  effects are
the lowering of the critical temperature and raising of the critical volume
to some extent.
The coexistence lines and the isothermal 
spinodal lines are shown by the dashed and the
dotted lines, respectively.
The critical temperatures and the critical volumes for different 
parameter sets are shown in Table I. 
The finite size effects as well as the
Coulomb interaction tend to reduce the critical temperature.
To isolate the Coulomb effect, the critical parameters for
$^{150}Sm$ are also displayed in 
Table I switching off the Coulomb 
interaction. It is seen that the Coulomb interaction  
lowers  $T_c$ for $^{150}Sm$
by about 1 MeV. 
The nature of EOS shows
that  it is possible to have a liquid-gas phase transition
in a finite nuclear system below the critical temperature $T_c$
if it is prepared suitably in  thermodynamic equilibrium.

To make a quantitative comparison of the
results  in the RTF theory with those obtained in the
non-relativistic Thomas-Fermi (NRTF) framework \cite{de1},  
the NRTF results for the critical temperatures 
and the critical volumes for
nuclear matter and for the nucleus $^{150}Sm$  are also given  in
Table I. The NRTF calculations were performed with a modified
Seyler-Blanchard (SBM)
effective interaction which gives $K_\infty\approx 240 $ MeV. This
lies in between those obtained with NL1 and NL3 parameter sets. 
However, it is seen that the critical parameters  for the SBM 
calculations are in between NL3 and NLSH for infinite nuclear matter
and close to those obtained with the NL3 parameter set for finite
systems. The nature of 
the EOS and the critical parameters are  controlled by the single-particle
potential and the effective mass. To compare these quantities in the RTF
and in the NRTF models, we display in Fig. 4 the nucleon
single-particle potential  as a function of the nuclear density 
(scaled with the normal nuclear density $\rho_0$)
for symmetric
nuclear matter as given in  the two models. The single particle
potential in the RTF is taken to be ${\cal V}+{\cal S}$
as given by Eqs.(7) and (8). The 
corresponding effective masses are shown in Fig. 5. 
It is seen that the SBM single-particle
potential is very close to  that obtained with the 
NL1 parameter set at lower densities which evolves 
towards that generated with  the NL3 parameter set with increasing density. 
The effective mass in the NRTF model is only a few percent lower 
at low densities and near the normal nuclear matter density it 
becomes somewhat higher
compared to those  obtained in the RTF model.
The single particle potential for protons as a function of  
distance from the centre
of  the  nucleus ($^{150}Sm$) in the two models corresponding to the
ground state and at a temperature $T = 8$ MeV are shown in the
top and the bottom panels of  Fig. 6, respectively.
The neutron single-particle potentials (not shown) have very similar
behavior. It is seen that  the potentials obtained
in the NRTF model and  those obtained  with
the different parameter sets in the RTF models are quite close. The effective
mass for finite nuclei in the two models is consistent with that
shown in Fig.5, i.e., the NRTF model yields effective mass which is
a little higher at the center and lower at the surface compared to the RTF
ones.  The above results  clearly demonstrate
the closeness of the single-particle potential and the
effective mass  in the non-relativistic and in the relativistic
framework. So it is natural to expect that the EOS and the related thermodynamic
properties would not be very different.

\subsection{ The caloric curve}
We have remarked before  that the density and hence
the observables depend on the volume in which  the nucleus at finite
temperature is confined. The calculation of the excitation energy as a function
of temperature (the caloric curve) is thus volume dependent.
In the experimental situation with energetic heavy ion collision,
it is generally assumed that the hot nuclear system prepared after the
collision expands substantially  beyond its normal size ($\sim 4 - 8$ times $V_0$)
and then undergoes fragmentation due to density instabilities. Guided by the
practice that many theoretical calculations for heavy ion collisions are
done by imposing that  thermalization occurs in a freeze-out volume, we
fix a volume and then find the caloric curve. The freeze-out
volume $V$ is determined by  $r_{max}$ occuring in Eqs. (\ref{epart}) -
(\ref{entropy}) as $V/V_0 = (1/A)(r_{max}/r_0)^3$. Here $r_0$ is the
radius parameter corresponding to the normal nuclear volume $V_0$; 
it is taken to be 1.2 fm.

In order to see the signature of liquid-gas phase transition, it is 
evident that the freeze-out volume is to  be chosen  beyond the
critical volume $V_c$. It  is seen from Table I that for the nuclei
considered, $V_c\sim 5 - 6$ times $V_0$. We have fixed
the freeze-out volume as $8V_0$ for our calculations. In Fig. 7,
the caloric curves for the nuclei   $^{40}Ca$, $^{109}Ag$ 
and $^{150}Sm$  are displayed  with  the three different 
parameter sets. The caloric curves  are seen to be nearly
independent of the parameter sets. At lower temperatures , the excitation energy
per particle $\epsilon^*$ increases quadratically with temperature
similar to that  in a Fermi gas; for $T$  between 5 - 10 MeV, the caloric curve
exhibits a shoulder and beyond that
with a kink $\epsilon^*$ rises linearly with temperature as
observed for  a classical gas.
In the bottom panel of Fig. 7  the data for experimental caloric curves
for the ALADIN and the EOS collaboration  are also shown.
It is noted that the calculated caloric curves 
show  shoulders at temperatures significantly higher than
those obtained from experiments. This may be attributed to
(i) the neglect of fluctuations in the theory which is expected to 
play important role near the transition temperature and  (ii)
the neglect of collective flow which lowers the
transition temperature appreciably  as seen in the non-relativistic calculations
\cite{de1}. 
The specific heat $C_v$ defined by Eq. (\ref{cv})
for the three parameter sets are shown in Figs. 8 and 9 for the three
systems studied.  Except for the
top pannel in Fig. 8 ,  the calculations reported are 
done at  the freeze-out volume $8V_0$. The
heat capacity shows a sharp  peak signalling the
liquid-gas phase transition,  the peaks occurring
at those temperatures where the caloric curve exhibits a kink. It is found that
the  harder the EOS the larger the transition temperature ($T_p$) and 
the sharper the peak.
In the top panel of Fig. 8 , the specific heat
for $^{40}Ca$ with a freeze-out volume 4$V_0$ is displayed. Instead of
a sharp peak, it shows a broad bump at $T\sim 10$ MeV.
Here the freeze-out volume is less than the critical volume, 
we do not associate this bump with a liquid-gas phase transition;
it possibly signals a precursor  to the transition.

The evolution of the density distributions 
for the three systems around the phase transition  temperature $T_p$ are
displayed in Figs. 10 and 11. The calculations are done in a freeze-out volume
8$V_0$. In Fig. 10, the  density distribution for 
$^{40}Ca$ with the parameter sets
NL3 (top panel) and NLSH (bottom panel) are shown.
At low temperatures,  the density is  more like a  Woods-Saxon profile; with
increasing temperature, the central density  depletes and a long tail spreading to
the boundary develops as is shown by the dashed line at $T$ = 0.9$T_p$. With
further increase in temperature  to $T_p$ and a little beyond, the density 
distribution tends to be uniform as is evident from the solid line ($T$ = $T_p$)
and the  dotted line ($T = 1.1T_p$). It is found that  the
evolution of density with temperature is nearly independent 
of  the parameter sets and therefore for the nuclei
$^{109}Ag$ and $^{150}Sm$ calculations with only NL3 parameter sets
are shown in Fig.11. For these nuclei too the evolution of
density is very similar to that in $^{40}Ca$. The  
rapid change  in the density distribution towards a uniform one
as the temperature  approaches $T_p$ is a further indication
of a liquid-gas phase transition in these finite systems.

For symmetric nuclear matter, at a fixed temperature, the pressure
remains constant over the whole coexistence region. The Gibbs free
energy per particle $g$ then shows a kink at the transition temperature when
the pressure is held constant. Its derivative with respect to 
temperature, the entropy function, then shows a discontinuity there.
For asymmetric nuclear matter, it has been shown in Ref.\cite{mul}
that the liquid-gas phase transition is second-order, the continuous
transition becoming more conspicuous  with increasing asymmetry.
There the pressure is not constant but 
shows a negative slope in the coexistence region 
for an isotherm in the $P-V$ plane. Thus
at constant pressure, the end points of the coexistence  line
are at different temperatures, then the kink in the Gibbs free
energy $g$ disappears and the entropy function becomes continuous
with kinks at the end points of the transition. The liquid-gas phase
transition thus occurs over a finite temperature interval. For
finite nuclei, the exact calculation of the thermodynamic functions 
in the coexistence region is nontrivial and still not known, but
as explained earlier in the beginning of this section, we expect
the pressure in this region to remain close to a constant; it is
then obvious that the entropy at constant pressure would show
a discontinuity at a transition temperature as shown in Fig.12
indicating a first-order phase transition. Precise statement 
about the order of the phase transition in finite nuclei can
be made only if the exact calculations of the partition functions
and hence the thermodynamic variables are rendered posssible. 
The present mean-field theory shows that the liquid-gas phase 
transition in finite nuclei has characteristics closer to
a first-order transition. This is in consonance with the results
obtained in the lattice gas model \cite{pan}. No definite conclusion
can however be reached from the experimental data; the GSI data
\cite{poc} indicate a first-order phase transition whereas the
analyses of the EOS data \cite{ell} show that the transition may
be second-order.

\section {Conclusions}

The relativistic mean-field theory which  has been
very  successful in describing the ground state properties
of nuclear systems has been applied for the first time in evaluating
the  equation of state of finite nuclei in this paper.
We have  resorted  to a  relativistic self-consistent
Thomas-Fermi theory  with the non-linear $\sigma-\omega-\rho$
version of the  Lagrangian. It is found that  the results do not differ
qualitatively from those obtained  in a non-relativistic  approach. 
This is due to the fact that the single particle potential and the effective 
mass which control
the relevant observables are very similar. The
calculations have been done with three parameter sets 
with very different  nuclear incompressibilities,
still the EOS look nearly the same and the critical 
parameters are not too different.
The critical parameters
for finite nuclei are appreciably different from those
of the infinite system. The specific heat  $C_v$ calculated 
from the caloric curve shows peaked
structures signalling liquid-gas phase transition  in the
nuclei studied. The  near uniformity of 
the density distribution  as the system approaches  the transition 
temperature confirms this further. Analysis of the thermodynamical
quantities indicates that this liquid-gas phase transition in the finite 
nuclei is more compatible with a first-order one.

\newpage
\begin{center}
\begin{table}
\caption{Critical temperature and critical volume for a few systems in the
RTF model with parameter sets NL1, NL3 and NLSH; and in the NRTF model
with the SBM interaction.}
\begin{tabular}{|c|c|c|c|c||c|c|c|c|}
%\hline
\multicolumn{1}{|c|}{Systems}&
\multicolumn{4}{c||}{$T_c$(MeV)}&
\multicolumn{4}{c|}{$V_c/V_0$}\\
\cline{2-9}
\multicolumn{1}{|c|}{}&
\multicolumn{1}{c|}{NL1}&
\multicolumn{1}{c|}{NL3}&
\multicolumn{1}{c|}{NLSH}&
\multicolumn{1}{c||}{SBM}&
\multicolumn{1}{c|}{NL1}&
\multicolumn{1}{c|}{NL3}&
\multicolumn{1}{c|}{NLSH}&
\multicolumn{1}{c|}{SBM}\\
\hline
Sym. NM& 13.4& 14.2& 15.4& 14.5& 3.9& 3.3& 2.9& 2.8\\
\hline
Asy. NM& 12.7& 13.6& 14.7& 14.1& 4.1& 3.4& 3.0& 2.9\\
(X = 0.2)&  &  &  &  &  & &  & \\
\hline
$^{40}Ca$  &11.1& 11.6& 12.4& --& 6.7 &6.5 &5.8&-- \\
\hline
$^{150}Sm$ & 11.4&  12.0& 13.0& 11.8& 5.4&5.0  &4.4& 4.9  \\
(With Coulomb)&  &  &  &  &  &  &  & \\
\hline
$^{150}Sm$& 12.3 &  12.9&14.0  & 12.5& 6.4  &6.0  &5.2&6.2  \\
(No Coulomb)&  &  &  &  &  & &  & \\
%\hline
\end{tabular}
\end{table}
\end{center}
\newpage
\noindent {\bf  Figure  Captions}

\begin{itemize}
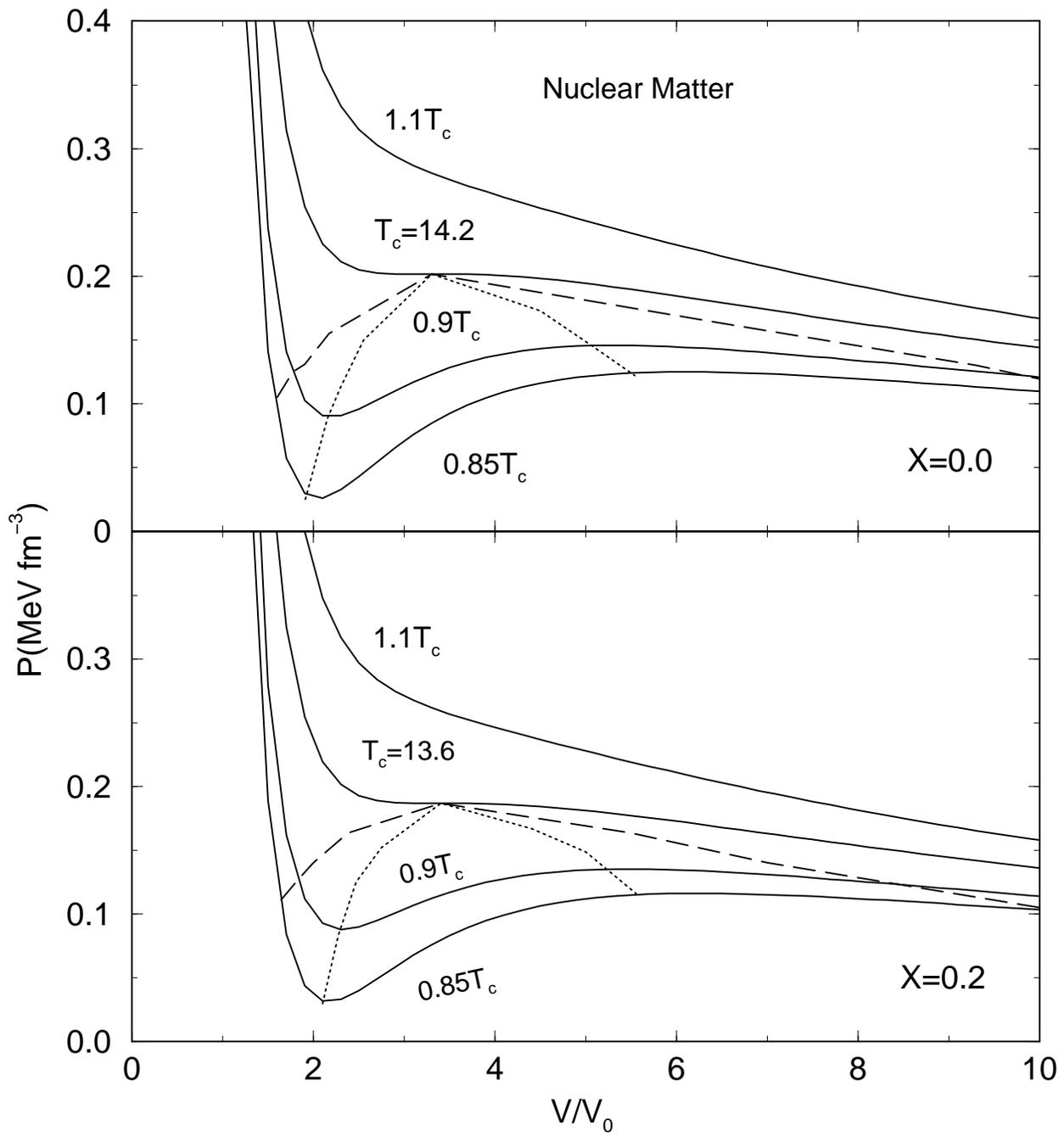
\item[Fig. 1] The equation of state  for symmetric nuclear matter
(top panel) and of asymmetric nuclear matter (bottom panel) with NL3 parameter
set. The temperatures (in MeV)  for the isotherms are as marked
in the figure. The dotted lines are the spinodals and the dashed lines
are the coexistence curves.

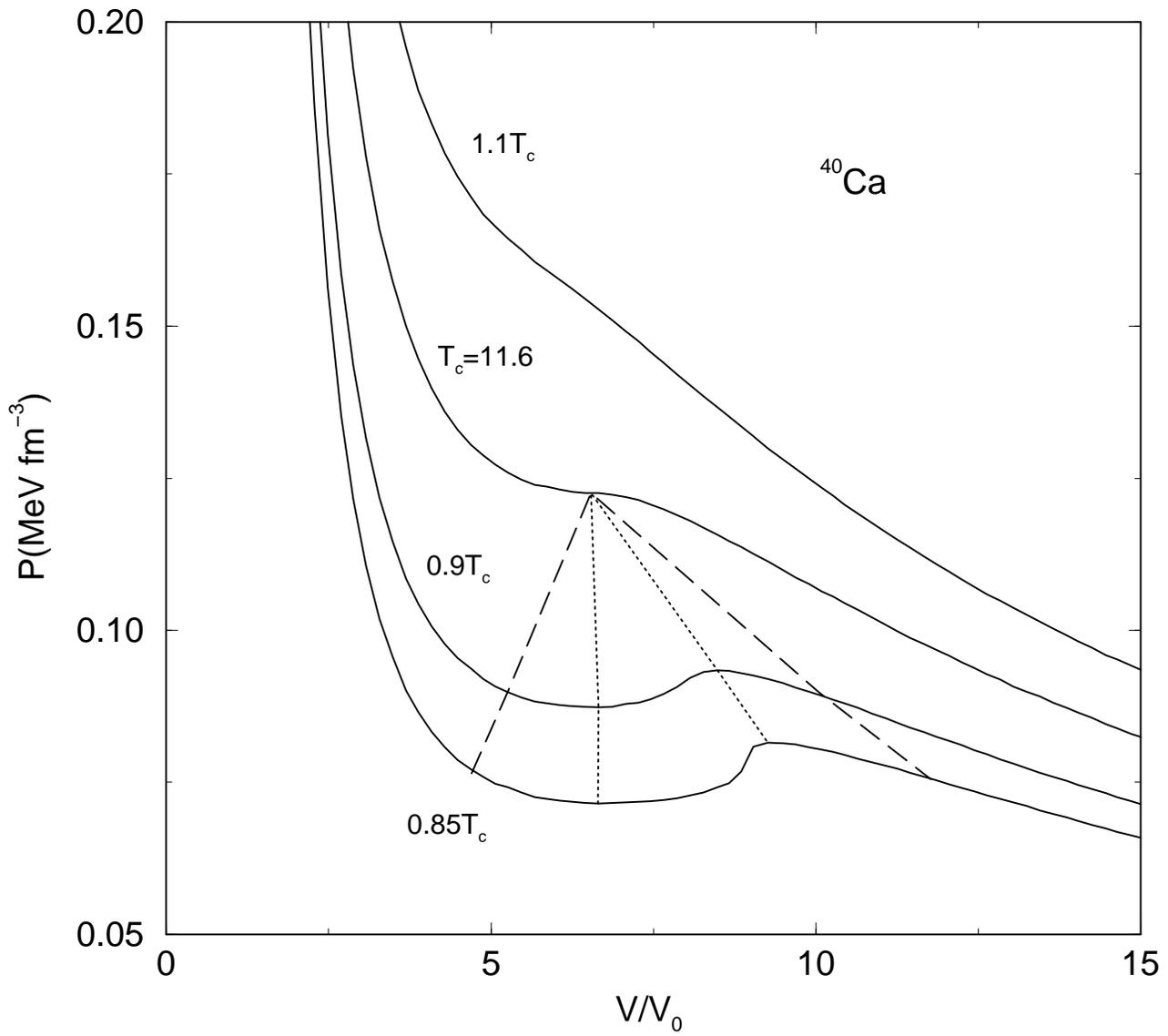
\item[Fig. 2] The equation of state for the nucleus $^{40}Ca$
for the NL3 parameter set. The different 
notations used have the same meaning as in Fig. 1.

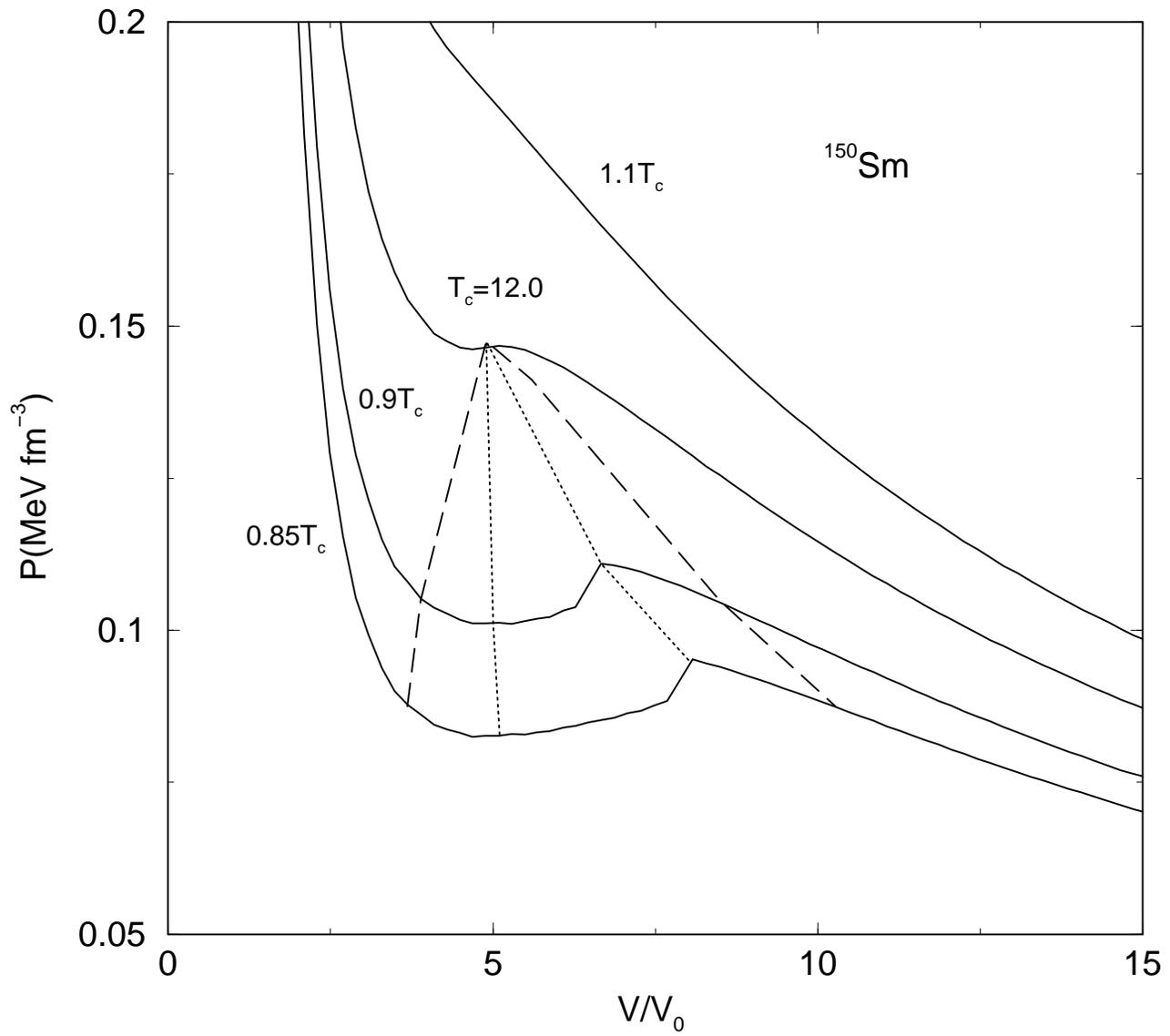
\item[Fig. 3] Same as in Fig. 2 but for the nucleus $^{150}Sm$.

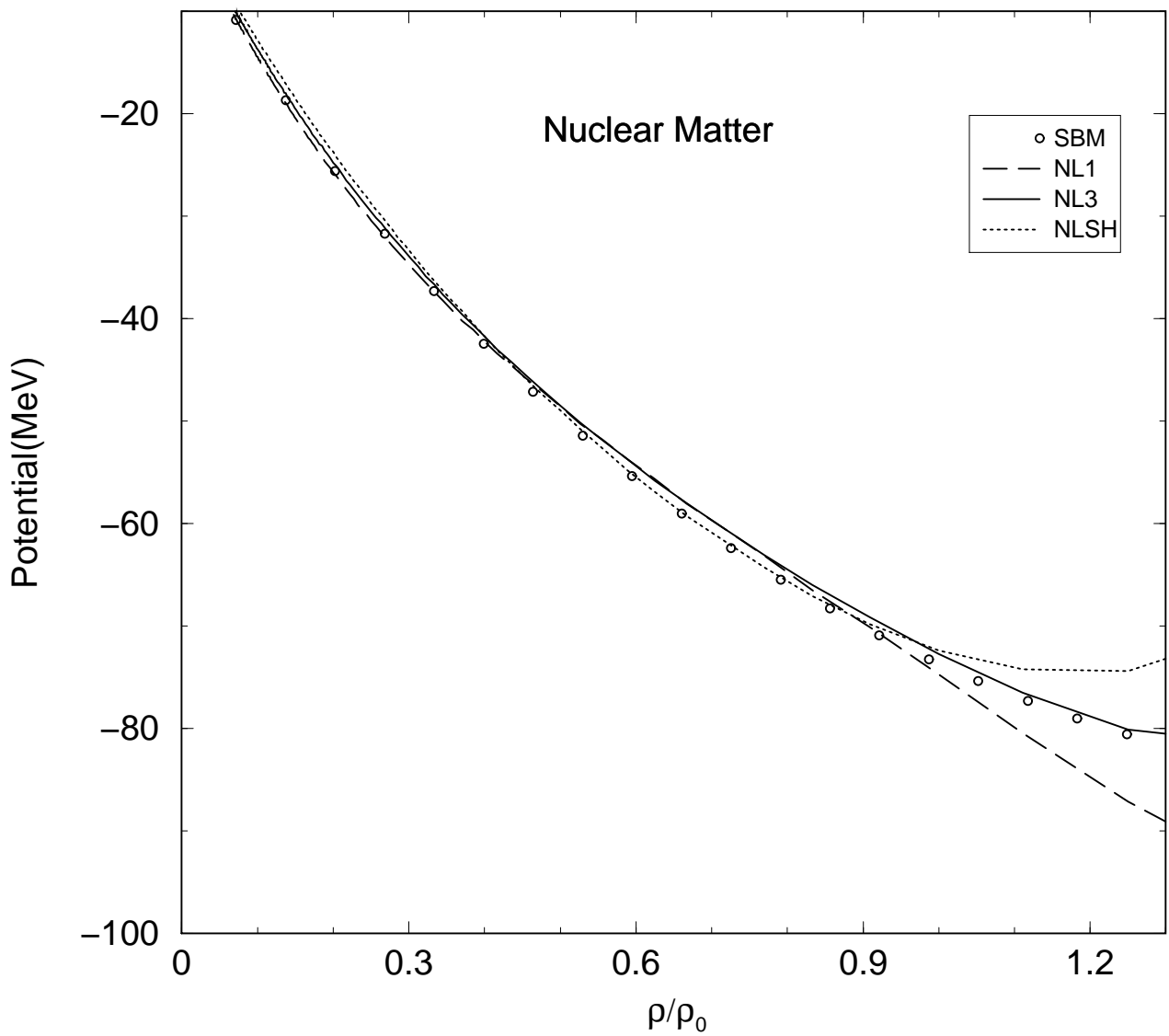
\item[Fig. 4] Plot for the relativistic and non-relativistic mean field
potentials as a function of density for symmetric nuclear 
matter at zero-temperature.

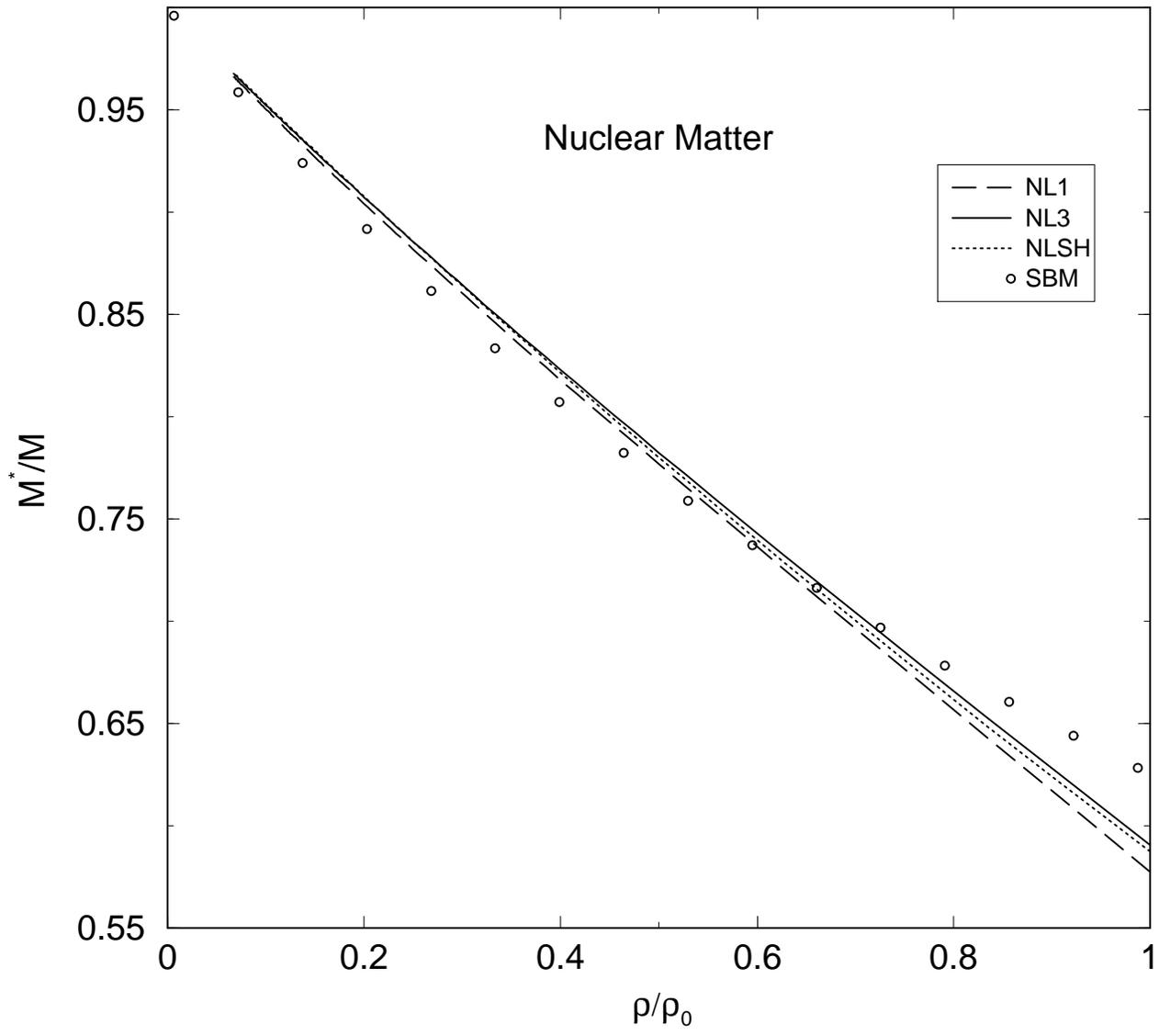
\item[Fig. 5] Density dependence of nucleon effective mass for symmetric
nuclear matter at zero temperature calculated within the relativistic and
non-relativistic framework.

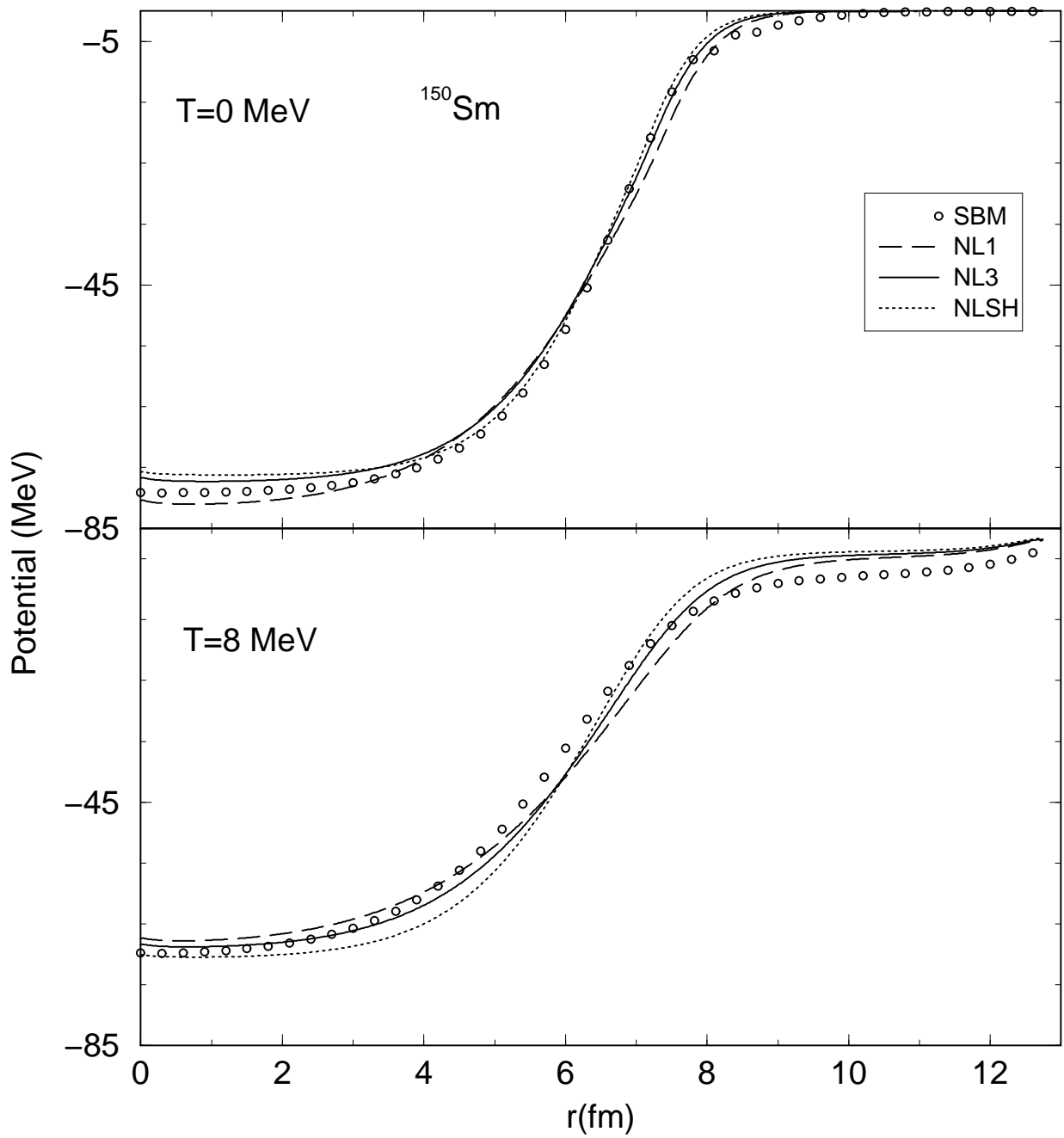
\item[Fig. 6] Relativistic and non-relativistic mean field potentials 
for protons at temperatures $T=0$ MeV (top panel) and $T=8$ MeV 
(bottom panel) for the nucleus $^{150}Sm$.

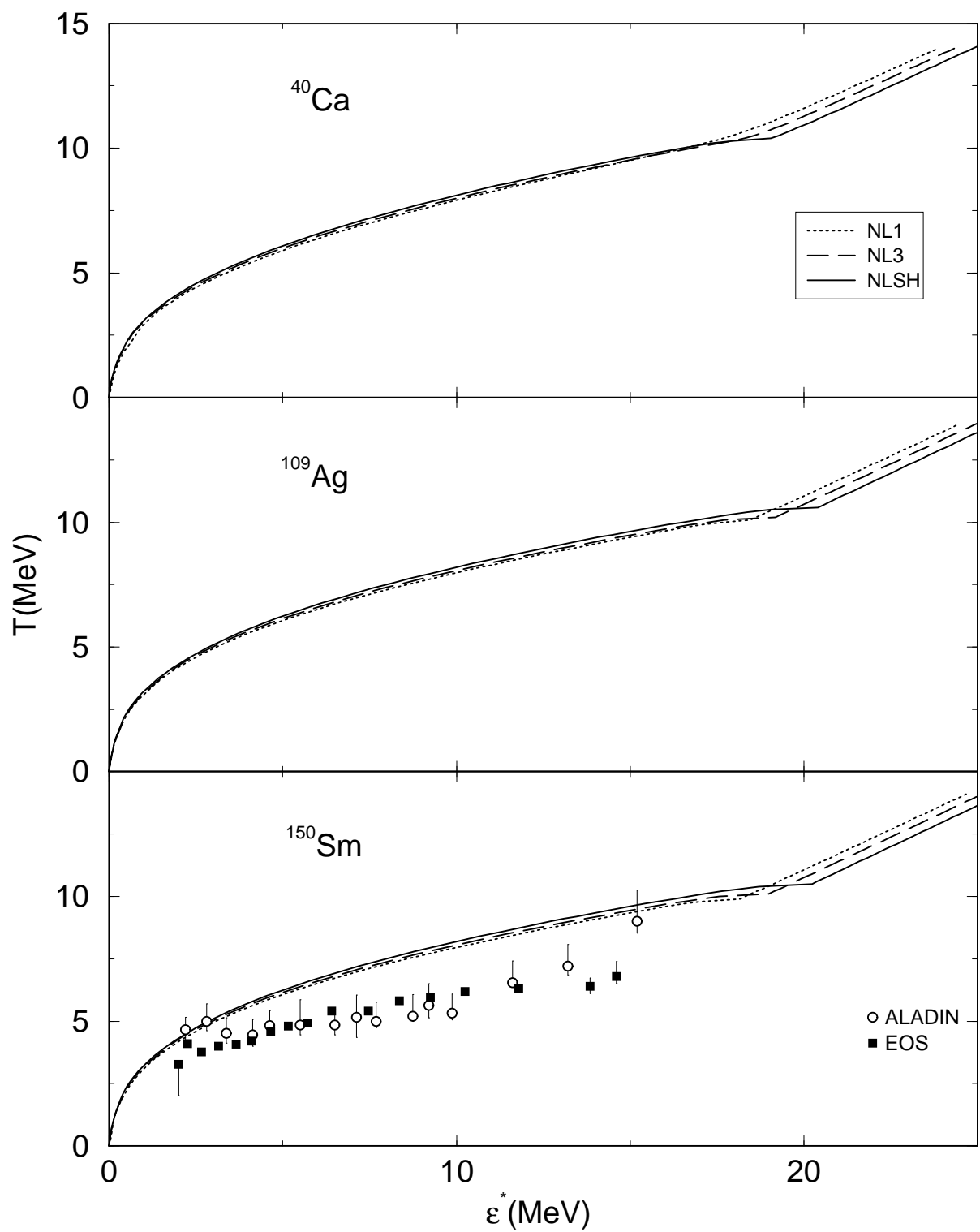
\item[Fig. 7] Caloric curves for the systems $^{40}Ca$, 
$^{109}Ag$  and $^{150}Sm$.
Open circles and filled squares  in the bottom panel represent
the experimental data for ALADIN and EOS collaborations.

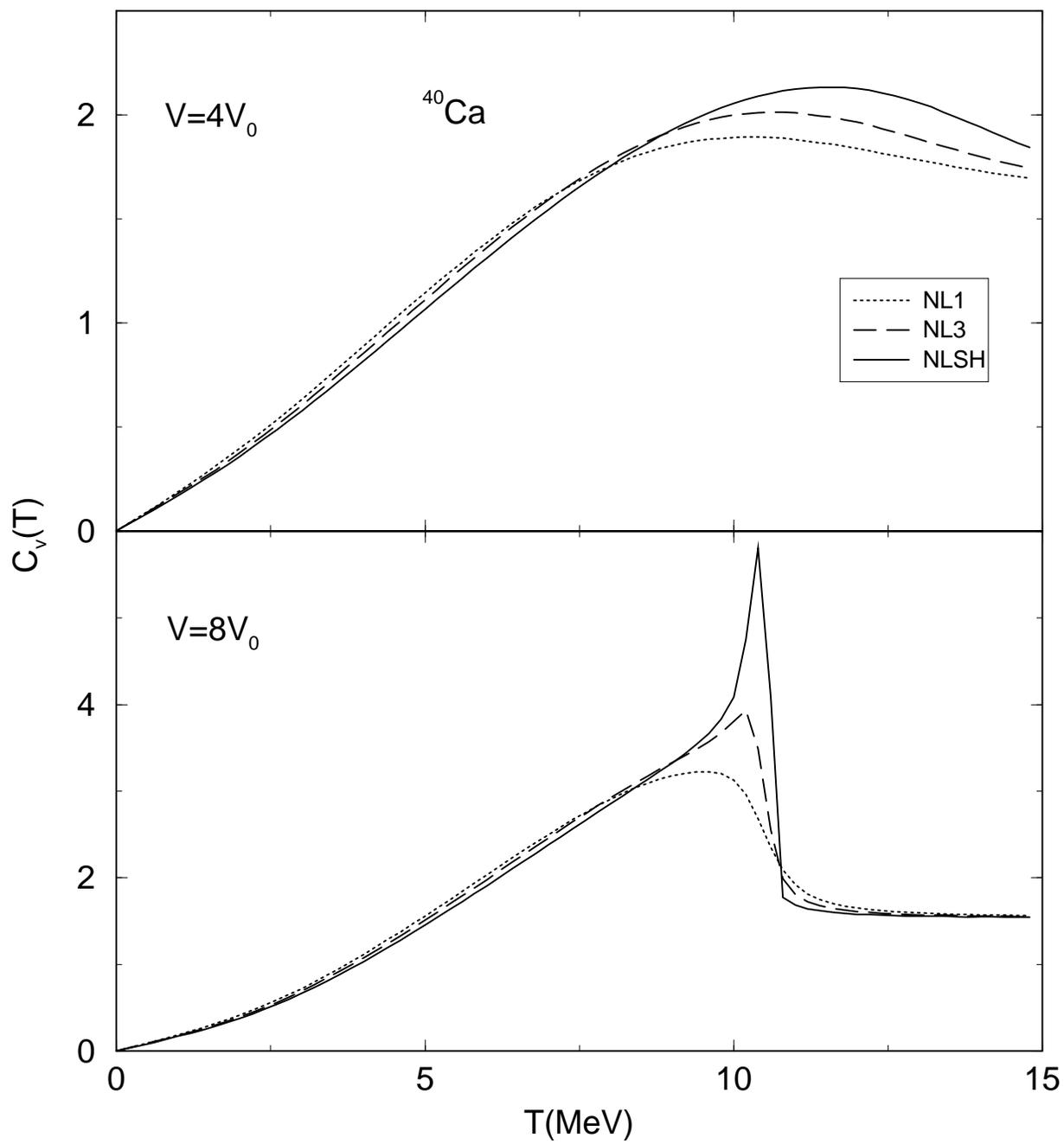
\item[Fig. 8] The specific heat for $^{40}Ca$ at
freeze-out volume equal to $4V_0$ (upper panel)  and $8V_0$ (lower 
panel) with different parameter sets as labelled.

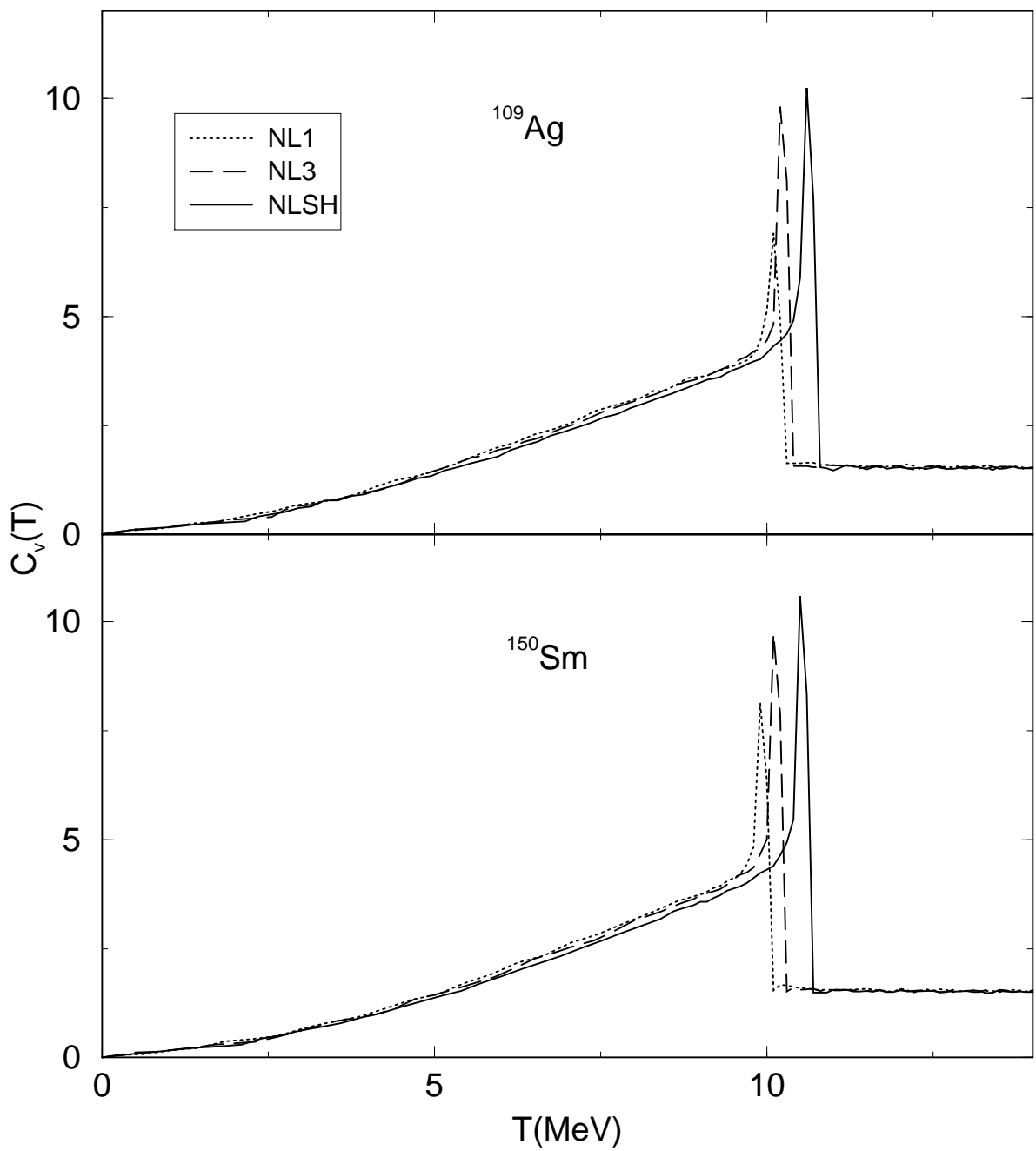
\item[Fig. 9] The specific heat for the nucleus $^{109}Ag$ (upper panel)  and
$^{150}Sm$ (lower panel) at the freeze-out volume $8V_0$ 
with different parameter sets.

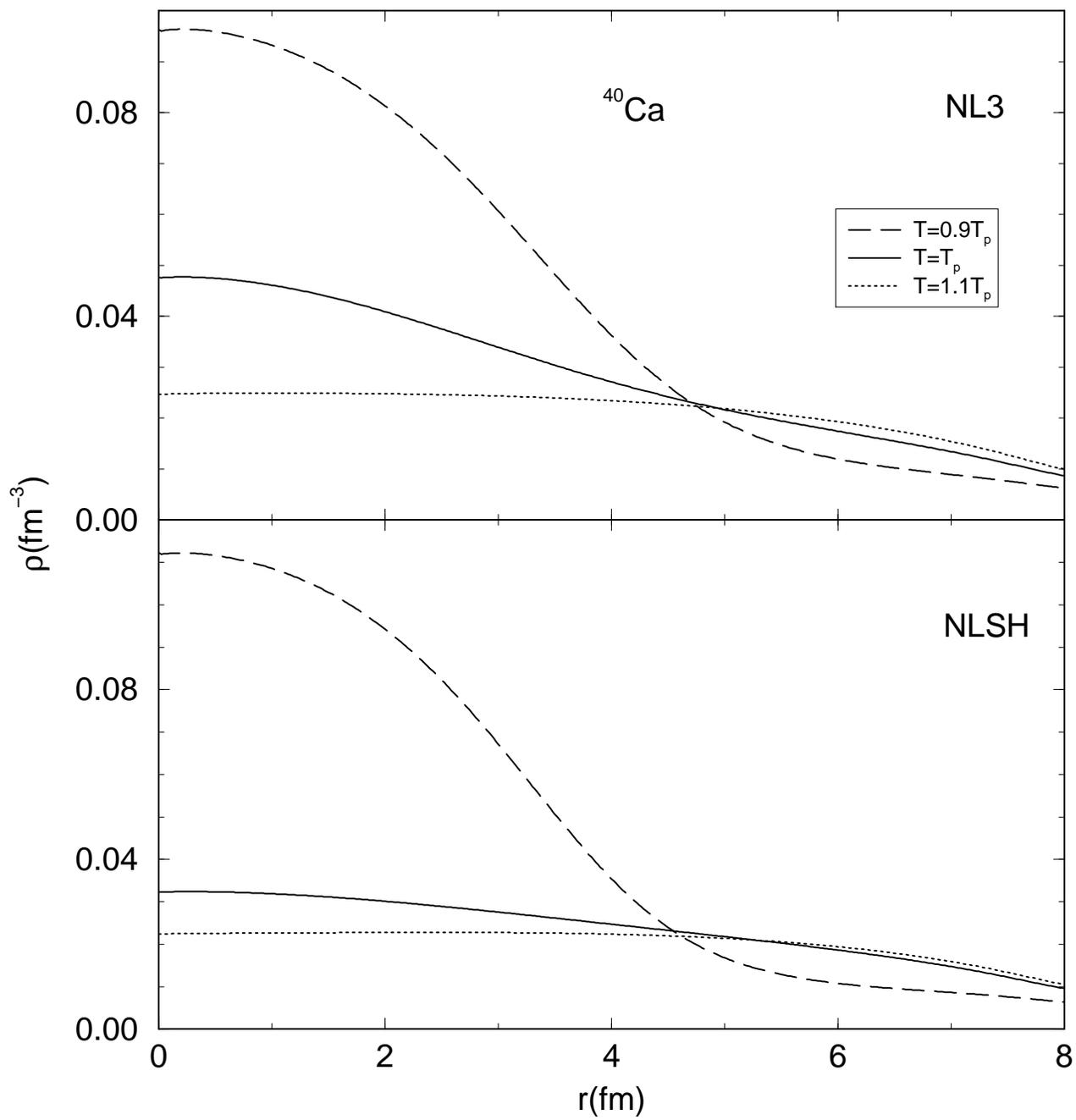
\item[Fig. 10] Density distributions  around transition 
temperatures $T_p=10.1$ and 10.3 MeV at a freeze-out
volume $8V_0$ for $^{40}Ca$
with parameter sets NL3 (upper panel) and NLSH (lower panel), respectively.

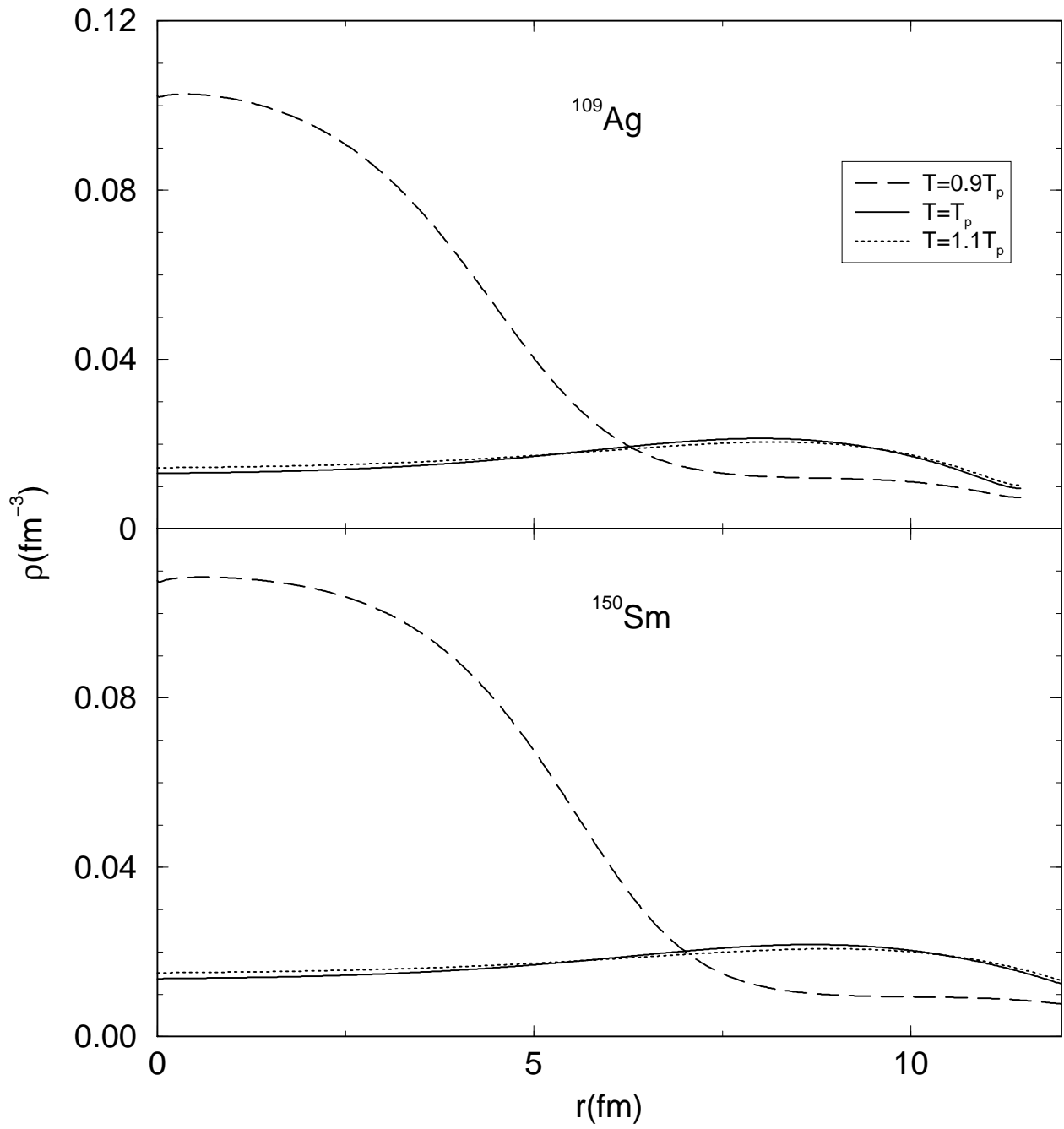
\item[Fig. 11] Density distribution around the transition temperatures
$T_p=10.2$ and 10.3 MeV
for $^{109}Ag$ (upper panel) and $^{150}Sm$ (lower panel), respectively 
with NL3 parameter set at a freeze-out volume $8V_0$.
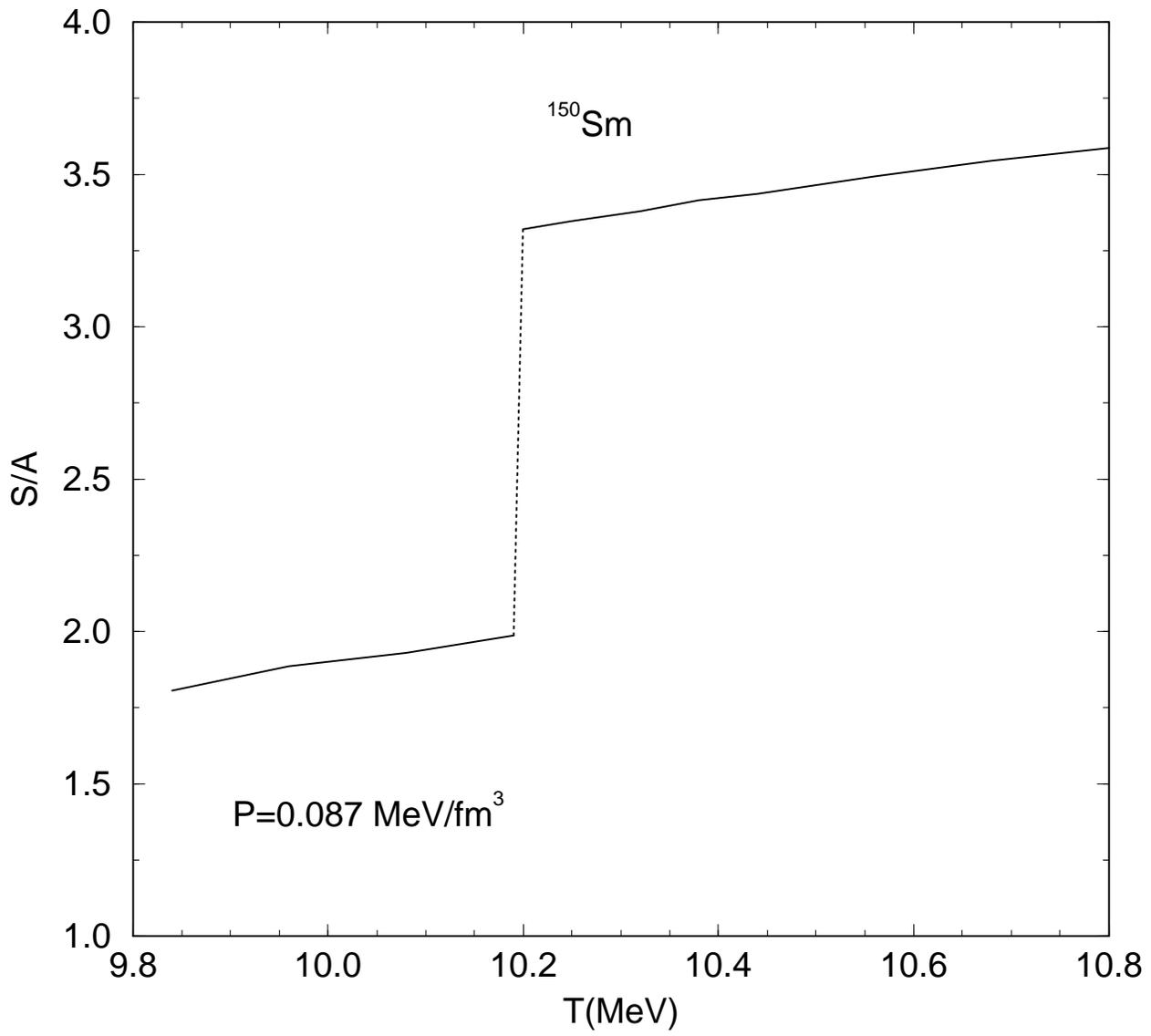
\item[Fig. 12] The temperature evolution of entropy per particle for 
the system $^{150}Sm$ at a constant pressure ($P$ =0.087 MeV/fm$^3$).
The dashed line shows the discontinuity at the transition temperature.
\end{itemize}

\begin{thebibliography}{999}
\bibitem{poc} J. Pochodzalla et al, Phys. Rev. Lett. {\bf 75}, 1040 (1995).
\bibitem{hau} J. Hauger et al, Phys. Rev. Lett. {\bf 77}, 235 (1996).
\bibitem{cib} J. Cibor et al, Phys. Lett. {\bf B473}, 29 (2000)
\bibitem{jaq} H. Jaqaman, A. Z. Mekjian and L. Zamick, Phys. Rev.
{\bf C27}, 2782 (1983); {\bf C29}, 2067 (1984).
\bibitem{ban} D. Bandyopadhyay, C. Samanta, S. K. Samaddar and J. N. De,
Nucl. Phys. {\bf A511}, 1 (1990).
\bibitem{ser} B. D.Serot and J. D. Walecka, Adv. in Nucl. Phys. {\bf 16},
1 (1986).
\bibitem{mul} H. M$\ddot u$ller and B. D. Serot, Phys. Rev. {\bf C52},
2072 (1995).
\bibitem{de1} J. N. De, B. K. Agrawal and S. K. Samaddar, Phys. Rev. {\bf C59},
R1 (1999).
\bibitem{de2} J. N. De, S. Dasgupta, S. Shlomo and S. K. Samaddar,
Phys. Rev. {\bf C55}, R1641 (1997).
\bibitem{gam} Y. K. Gambhir, P. Ring and A. Thimet, Ann. Phys. (N.Y),
{\bf 198}, 132 (1990).
\bibitem{lal} G. A. Lalazissis, D. Vretenar, W. Poeschl and P. Ring,
Nucl. Phys. {\bf A632}, 363 (1998).
\bibitem{vre} D. Vretenar, G. A. Lalazissis and P. Ring, Phys. Rev. Lett.
{\bf 82}, 4595 (1999).
\bibitem{bog} J. Boguta and A. R. Bodmer, Nucl. Phys. {\bf A292}, 413 (1977).
\bibitem{bra} M. Brack, C. Guet and H. B. Hakansson, Phys. Rep. {\bf 123},
275 (1985).
\bibitem{de3} J. N. De, N. Rudra, Subrata Pal and S. K. Samaddar,
Phys. Rev. {\bf C53}, 780 (1996).
\bibitem{lal2} G. A. Lalazissis, J. K$\ddot o$nig and P. Ring, 
Phys. Rev. {\bf C55}, 540 (1997).
\bibitem{bar} M. Barranco and J. R. Buchler, Phys. Rev. {\bf C22},
1729 (1980).
\bibitem{pat} R. K. Pathria, Statistical Mechanics, Pergamon Press,
p.377 (1985).
\bibitem{pan} J. Pan, S. Dasgupta and M. Grant, Phys. Rev. Lett.
{\bf 80}, 1182 (1998).
\bibitem{ell} J. B. Elliot et al, Phys. Lett. {\bf B381}, 35 (1996).
\end{thebibliography}
\end{document}